\documentclass[final,5p,twocolumn]{elsarticle}

\usepackage{graphicx}
\usepackage[inline]{enumitem}
\usepackage{amsmath, amssymb, amsfonts}
\usepackage{array}
\usepackage{hyperref}
\usepackage{xcolor}
\usepackage{multirow, multicol}
\usepackage[table,dvipsnames]{xcolor}
\usepackage{tabularx}
\usepackage{tabularray}
\usepackage{booktabs}
\usepackage{balance}
\usepackage{tabularray}

\newcolumntype{Y}{>{\centering\arraybackslash}X}
\newcolumntype{Z}{>{\raggedright\arraybackslash}X}

\newcounter{inlineenum}
\renewcommand{\theinlineenum}{\roman{inlineenum}}
\newenvironment{inlineenum}
  {\unskip\ignorespaces\setcounter{inlineenum}{0}
   \renewcommand{\item}{\refstepcounter{inlineenum}{\textbf{\textit{\theinlineenum})~}}}}
  {\ignorespacesafterend}

\usepackage{makecell}
\newcolumntype{x}[1]{>{\centering\arraybackslash}p{#1}}
\usepackage{tikz}

\usepackage{array,longtable,booktabs,ragged2e,makecell}
\newcolumntype{L}[1]{>{\RaggedRight\arraybackslash}m{#1}}
\newcolumntype{C}[1]{>{\Centering\arraybackslash}m{#1}}
\renewcommand{\arraystretch}{1.15}
\title{Evidence-Driven Decision Support for AI Model Selection in \\Research Software Engineering}

\author[1]{Alireza Joonbakhsh}
\author[1]{Alireza Rostami}
\author[1]{AmirMohammad Kamalinia}
\author[1]{Ali Nazeri}
\author[1]{Farshad Khunjush}
\author[2]{Bedir~Tekinerdogan}
\author[2]{Siamak Farshidi \corref{cor1} \fnref{fn1}}
\ead{siamak.farshidi@wur.nl}

\cortext[cor1]{Corresponding author}
\fntext[fn1]{Email: \href{mailto:siamak.farshidi@wur.nl}{\texttt{siamak.farshidi@wur.nl}}}

\affiliation[1]{
    organization={Department of Computer Science and Engineering, Shiraz University},
    city={Shiraz},
    country={Iran}
}
\affiliation[2]{
    organization={Wageningen University \& Research},
    city={Wageningen},
    country={The Netherlands}
}

\begin{document}

\begin{abstract}
\textbf{Context:} The rapid proliferation of artificial intelligence (AI) models and methods has created new challenges for research software engineers (RSEs) and researchers who must select, integrate, and maintain appropriate models within complex research workflows. This selection process is often ad hoc, relying on fragmented metadata and individual expertise, which can compromise reproducibility, transparency, and research software quality.

\textbf{Objective:} This work aims to provide a structured and evidence-driven approach to assist research software engineers and practitioners in selecting AI models that best align with their technical and contextual requirements.

\textbf{Method:} The study conceptualizes AI model selection as a \textit{Multi-Criteria Decision-Making} (MCDM) problem and introduces an evidence-based decision-support framework that integrates automated data collection pipelines, a structured knowledge graph, and MCDM principles. Following the Design Science Research (DSR) methodology, the framework, referred to as \textsc{ModelSelect}, is empirically validated through 50 real-world case studies and comparative experiments with leading generative AI systems.

\textbf{Results:} The evaluation demonstrates that the framework yields reliable, interpretable, and reproducible recommendations that closely align with expert rationales. Across the case studies, \textsc{ModelSelect} achieved high coverage and rationale alignment in model and library recommendation tasks, performing comparably to generative AI assistants while providing greater traceability and consistency.

\textbf{Conclusion:} Modeling AI model selection as an MCDM problem provides a rigorous foundation for transparent and reproducible decision support in research software engineering. The proposed framework offers a scalable and explainable pathway for integrating empirical evidence into AI model recommendation processes and improving the quality of research software decision-making.

\end{abstract}


\maketitle


\section{Introduction}

In recent years, AI and its subset ML have enjoyed an unprecedented proliferation in terms of popularity and usage \cite{dwivedi2023evolution, hajkowicz2023artificial}, as AI/ML\footnote{For the sake of convenience, this work refers to AI/ML as AI, AI model as model, and AI model feature as feature from now on.} models and methods can be employed to carry out plenty of tasks that were previously performed manually, with superior performance for a variety of applications. As a result, AI has become an eye-catching option, sparking a significant amount of research on utilizing it in numerous domains and ecosystems \cite{rashid2024ai} such as healthcare \cite{habehh2021machine}, biology \cite{bhardwaj2022artificial}, chemistry \cite{baum2021artificial}, and finance \cite{hoang2023machine} for applications like classification, prediction \cite{habehh2021machine, bhardwaj2022artificial}, and decision-making \cite{dwivedi2023evolution}. This use of AI in other domains is often performed under the umbrella of \textit{data science}. This multidisciplinary field focuses on combining domain/business knowledge with AI and software development to extract knowledge from data through AI models and methods \cite{filipe2021datascience}.

\smallskip


The field of AI is rapidly growing, with new models and methods being developed and proposed at an accelerating pace \cite{tang2020pace, maslej2024artificialintelligenceindexreport}. For instance, the number of AI papers published monthly doubles approximately every 23 months \cite{krenn2023forecasting}. These models vary significantly in features such as architecture, size, input/output modality, and more \cite{khan2020survey, li2023architecture}. For instance, one study reviewed 150 variations of text classification using the deep learning model alone \cite{minaee2020deeplt}, each with its own technical features and qualitative strengths and shortcomings.

\smallskip


This complexity poses a significant barrier for Research Software Engineers (RSEs) tasked with developing AI software for research. RSEs typically come from domains outside of AI and often lack the expertise needed to navigate the AI landscape \cite{rani2025empirical}. They face the task of selecting the best model for their specific context, a decision that extends far beyond standard accuracy metrics. It involves considering a model's architecture \cite{Jiang2023Empirical}, interpretability \cite{benchaaben2025model}, and practical project requirements like data privacy, regulatory compliance, and deployment constraints \cite{Arpteg2018Software, Barr2024Review}.

\smallskip


To exacerbate the problem, AI development is costly, both in terms of time and quality \cite{hill2016aidevelopment}, and choosing inappropriate models may lead to wasted time, resources, and costs, or to poor solution performance \cite{ding2018modelselection}. Furthermore, model choice affects fairness, accuracy, and user trust of the said solution \cite{papenmeier2022usertrust}. This critical decision is often made without structured guidance, forcing practitioners to rely on a number of workarounds.

\smallskip

In practice, RSEs often rely on informal and fragmented sources to identify suitable AI models \cite{wang2020automodel, li2023metadatarep}, as model metadata is scattered across multiple sources such as documentations, benchmarks, forums, and papers \cite{Jiang2023Empirical}. Reviewing these metadata requires browsing code repositories such as GitHub \cite{amershi2019}, consulting package indexes such as PyPI, reviewing online discussions on forums such as Stack Overflow or Reddit, or following recommendations from blog posts, tutorials, and social networks. Although helpful, critical model features can be overlooked when browsing these sources manually, as they tend to emphasize popularity or recency over suitability for specific technical needs, such as hardware requirements and reproducibility \cite{Jiang2023Empirical}. As a result, model selection remains largely manual, subjective, and time-consuming.

\smallskip

Mitigation strategies discussed in the literature range from improving and standardizing model documentation and metadata \cite{tsay2022extracting, li2023metadatarep} to providing replication and reproducibility packages \cite{Wang2023Machine}, and integrating software engineering best practices into AI workflows \cite{serban2024swpractices, Correia2020Brazilian}. Collectively, the evidence suggests that model selection is not a straightforward matter of selecting the highest-ranking option on a leaderboard. It is a multidimensional decision-making process that requires structured, transparent, and context-aware support.

\smallskip

GenAI assistants offer a potential solution to AI model selection but come with significant limitations. First, they often lack contextual awareness \cite {farshidi2025pyselect}, failing to account for project-specific needs, such as resource limitations, data privacy, and deployment challenges \cite{Arpteg2018Software}. Second, they are susceptible to hallucination \cite{huang2025survey}, presenting users with outputs that seem plausible yet incorrect, nonfactual, or unrelated to the task \cite{spracklen2024we}. Finally, they are not explainable \cite{Zhao2024explainable}, meaning mechanisms behind their responses are not transparent, and their recommendations often reflect biases \cite{xue2023biasfairnesschatbotsoverview} that exist in datasets for their training data \cite{spracklen2024we}.

\smallskip

The challenges of model selection are also not thoroughly addressed when consulting with open-source model registries and large repositories like HuggingFace, where transparency, provenance tracking, and portability across environments remain inconsistent \cite{Jiang2023Empirical}.

\smallskip

This lack of structured guidance presents a significant barrier for RSEs seeking to adopt AI models effectively. A clear need exists for a decision support approach that integrates objective metrics and subjective feedback to provide context-sensitive recommendations for research software development.

\smallskip

Such a need aligns with the broader view that in software engineering, every stage of the process (regarding design, technology selection, and integration) is essentially a decision-making activity \cite{Fitzgerald2014ContinousSE,Ruhe2020SEdecision}. Because these decisions are often knowledge-intensive and prone to bias from experience or incomplete information \cite{KahnemanTversky2000}, structured methods are essential to ensure balanced outcomes. MCDM frameworks are well-suited for this, as they systematically evaluate alternatives using both quantitative and qualitative criteria \cite{Doumpos2013mcdmAI}. In AI model selection for research software, the complexity of such decisions is amplified by the rapidly evolving ecosystem of models, making MCDM a suitable foundation for model choice \cite{farshidi2020multicriteria}.

\smallskip


To this end, this study proposes a novel framework for structured AI model selection. Our framework guides RSEs to find models that match their specific intent, similar to an AI expert. The system can compare models based on both key technical features and qualitative user feedback.

\smallskip

DSR is well-suited for addressing such MCDM problems in software production, as it provides a systematic approach to selecting and validating artifacts (such as AI models) \cite{farshidi2020multicriteria}. Therefore, we employed a Design Science Research (DSR) methodology to develop and evaluate our decision model. Data was collected from public sources, including official documentation, package indexes, and community platforms. Our approach incorporates both an objective feature list and subjective community feedback. Additionally, the decision model integrates MCDM principles, combining model features with user sentiment analysis.

\smallskip

The decision model is validated using 50 carefully selected case studies from real-world applications of AI models. Each case study was selected from published research in reputable journals, with most including accompanying code repositories. The results highlight that integrating both the subjective and objective criteria enhances decision-making accuracy, helping RSEs identify models that align with their technical requirements. The findings also highlight the importance of a context-aware framework that can be adapted to other AI ecosystems beyond Python.

\smallskip

This work contributes directly to the engineering of GenAI-enabled software systems by designing and evaluating an evidence-driven decision-support framework. The system, \textsc{ModelSelect}, integrates GenAI components into key stages of the software engineering process, including metadata extraction, sentiment-based quality assessment, and model-feature mapping. In doing so, it demonstrates how GenAI can be systematically embedded within software development pipelines. Beyond improving automation and transparency, the framework represents a new class of GenAI-enabled engineering tools that enhance decision-making, reproducibility, and explainability in the construction of AI-based software systems.

\smallskip

To summarize, our main contributions are as follows:
\begin{itemize}
    \item A novel decision model for AI model selection that combines model features with user sentiment analysis, incorporating both objective metrics and subjective community feedback.
    \item A curated evaluation dataset of 50 case studies with annotated model choices and rationales.
    \item An empirical evaluation that comprises a validation against the case-study dataset and a comparative study with generative AI tools, demonstrating the system's effectiveness and usability.
    \item A rationale-alignment evaluation metric to assess recommendation validity.
\end{itemize}

\smallskip

\phantomsection
\label{introduction:prototype}

All resources required to reproduce and extend this work are openly available in a curated Mendeley Data repository \cite{modelselectmendeley}. This repository includes \begin{inlineenum}
    \item the complete source code for all automated pipelines described in Section~\ref{subsection:pipelines},
    \item data collected during pipeline execution, including library documentation and model metadata,
    \item experiment sheets reporting pipeline performance results,
    \item spreadsheets representing the literature study, and
    \item the evaluation case-study corpus with corresponding ground truth, system outputs, and analyses.
\end{inlineenum}
In addition, a prototype recommender system developed as part of this research is publicly accessible\footnote{A live prototype of the recommender system, \textit{MLSelect}, is available at \href{https://ai4rse.nl/MLModelSelection/search/}{\texttt{https://ai4rse.nl/MLModelSelection/search/}}, where users can explore AI model recommendations based on evidence-driven decision criteria.}.

\smallskip


The rest of the paper is organized as follows:
Section \ref{research_approach} formally defines the problem, outlines the research questions, and describes the methodologies employed to address them. Section \ref{modelselect_framework} introduces the \textsc{ModelSelect} framework, and its components. Section \ref{metadata} analyzes the data in our knowledge base. Section \ref{evaluation} presents the evaluation of the data collection pipelines and the holistic assessment of \textsc{ModelSelect} against GenAI tools. Section \ref{discussion} discusses the results, limitations, and implications of the study. Section \ref{related_work} situates our approach within related work. Finally, Section \ref{conclusion} concludes the paper and outlines future work.


\begin{figure*}[ht!]
    \centering
    \includegraphics[width=\linewidth]{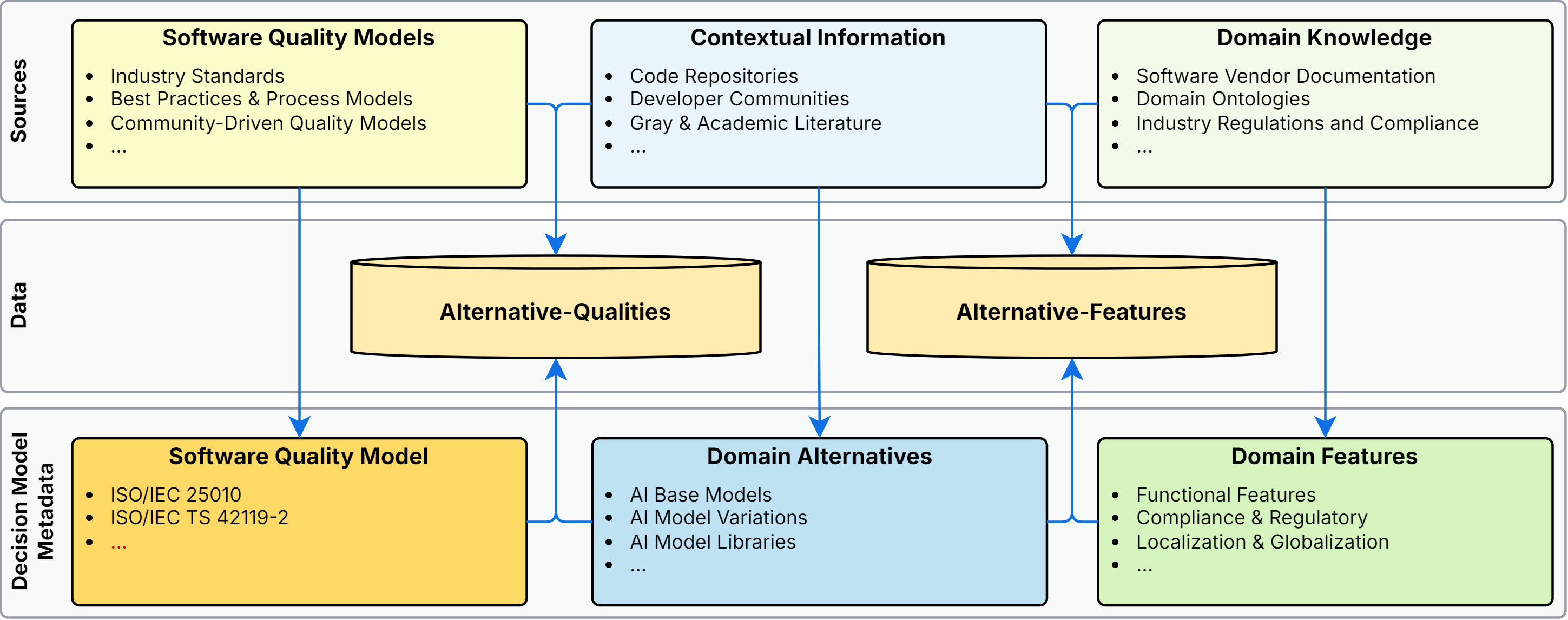}
    \caption{An adaptation of the decision model framework \cite{farshidi2020multicriteria}, extended with a data collection perspective that enables automated integration pipelines. This structure supports systematic, evidence-based evaluation of AI-based models, their variations, and libraries.}
    \label{fig:framework}
\end{figure*}

\section{Research Approach}
\label{research_approach}


This study employs a structured research approach to examine AI model selection in research software engineering, with a focus on open repositories and scholarly use contexts. Following the DSR methodology \cite{hevner2004design, peffers2007design}, our work incorporates artifact-oriented engineering with empirical inquiry. Our goal is to conceptualize, implement, and assess a theoretical MCDM framework for data collection, integration, and AI model recommendation. To realize this framework, we develop \textsc{ModelSelect}, a data-driven decision-support system that assists RSEs and practitioners in selecting AI models tailored to project-specific requirements and constraints.

\smallskip

The following subsections will discuss the terminology, problem formulation, research questions, and research methods applied in this study.

\subsection{Terminology}

To ground the discussion that follows, we clarify key terms used throughout this work.

\smallskip

An AI \textit{\textbf{model}} is a software artifact, such as an ML or Deep Learning (DL) architecture, along with its learnable coefficients trained on the data to perform tasks that include prediction, classification, generation, or decision support \cite{tsay2022extracting}.

\smallskip

An AI \textit{\textbf{library}}, refers to a reusable software library or package that implements AI models, learning algorithms, utilities, or supporting functions, thereby enabling developers to build, train, and integrate AI systems \cite{kiraly2021toolbox}. Within the Python ecosystem, such libraries are also referred to as \textit{packages}; therefore, this study uses the term \textit{library} to encompass both software libraries and Python (PyPI) packages. 

\smallskip

Furthermore, a Generative AI (\textit{\textbf{GenAI}}) refers to a class of AI systems capable of producing novel content (such as text, images, code, or other media) based on patterns learned from training data \cite{nist2024genai}. Throughout this work, we collectively refer to large language models (LLMs) and AI assistants as GenAI. 

\smallskip

An \textit{\textbf{RSE}} refers to data scientists and researchers responsible for developing AI software for research purposes.

\subsection{Problem Formulation}
\label{problem_formulation}

Technology selection in software engineering \cite{pressman2010software, farshidi2018multiple} involves identifying tools, components, or services that meet both functional \cite{wiegers2013software} and non-functional requirements \cite{chung2012non}. This process spans various domains, such as architectural patterns \cite{schmidt2013pattern, ArchitecturePatternsfarshidi2020capturing}, database systems \cite{Databasefarshidi2018decision}, cloud platforms \cite{Cloudfarshidi2018decision}, and third-party libraries \cite{farshidi2025pyselect, pace2023javascript}. The increasing number of alternatives, coupled with differences in quality, maturity, and ecosystem compatibility, makes the selection process more complex and prone to risk.

\smallskip

MCDM theory offers a structured approach to address this challenge by evaluating alternatives against predefined criteria \cite{thakkar2021multi}. Previous research has utilized MCDM theory to create a theoretical framework for modeling technology selection problems in software engineering \cite{farshidi2020multicriteria}. This framework has been instantiated in diverse areas, such as database management \cite{Databasefarshidi2018decision}, cloud service provisioning \cite{Cloudfarshidi2018decision}, blockchain platforms \cite{Blockchainfarshidi2020decision}, software architecture patterns \cite{ArchitecturePatternsfarshidi2020capturing, farshidi2020decision}, and Python package selection \cite{farshidi2025pyselect}. Additional applications have extended the framework to model-driven development \cite{farshidi2021model}, programming language ecosystems \cite{farshidi2021decision}, decentralized autonomous organizations \cite{DAObaninemeh2023decision}, business process modeling languages \cite{BPMLfarshidi2024business}, user intent modeling \cite{IntentModelfarshidi2024understanding}, and blockchain oracle platforms \cite{Oracleahmadjee2025decision}.

\smallskip

Previous implementations, however, faced a significant limitation, which was their reliance on manual data collection methods, such as expert interviews and literature reviews. Although effective in domains with fewer alternatives, manual approaches are impractical in dynamic fields like AI model selection. In these environments, models and their features evolve frequently, with relevant information scattered across diverse and heterogeneous sources. 

\smallskip

Despite these challenges, the AI model selection process remains well-structured. It aligns well with the MCDM paradigm, involving a clearly defined set of alternatives, domain-specific features, and a systematic mapping of project requirements to potential solutions. To this end, our proposed framework, illustrated in Figure \ref{fig:framework}, is instantiated to support the automated and evidence-based evaluation of AI models.

In this context, each model is represented as an alternative described by a vector of measurable features. The decision process identifies the alternatives that align the most closely with project-specific qualitative and functional requirements. Formally, let $Models = \{m_1, m_2, \dots, m_n\}$ denote the set of models, where each model is characterized by a set of attributes $Features = \{f_1, f_2, \dots, f_m\}$. The project requirements are represented as a subset $Requirements \subseteq Features$. The goal is to identify a solution set $Solutions \subseteq Models$ that maximizes alignment with the specified requirements. The mapping MCDM captures this relationship:
\begin{displaymath}
    Models \times Features \times Requirements \rightarrow Solutions  
\end{displaymath}

\subsection{Research Questions}

This study aims to address these challenges by formulating the following research questions:

\begin{enumerate}
    \smallskip
    \item[] \textbf{Main Research Question (MRQ):} How can RSEs be supported in selecting appropriate AI models according to their desired intent?
    
    \item[] \textbf{RQ1:} \label{RQ1} How can relevant data regarding AI models and their features be collected automatically in the context of generating the knowledge base?

    \item[] \textbf{RQ2:} \label{RQ2} How can the collected data be represented by a structured knowledge graph that supports a decision model for AI model selection?
    
    \item[] \textbf{RQ3:} \label{RQ3} How can the resulting decision model be evaluated to determine its effectiveness in supporting RSEs during AI model selection?

\end{enumerate}

\subsection{Research Methods}

\subsubsection{Design Science Research (DSR)}

This study employs a multiphase methodology grounded in the DSR paradigm, informed by the framework proposed by \cite{hevner2004design} and \cite{peffers2007design}. The aim is to design, implement, and evaluate a decision-support system for evidence-based AI model selection, realized through the \textsc{ModelSelect} software tool. 

\smallskip

The research activities align with the three guiding research questions and follow the DSR phases: \textit{Problem Identification} addresses the challenges RSEs face in AI model selection due to fragmented metadata and lack of scalable solutions; \textit{Objective Definition} establishes the goal of automating data collection, integration, and structuring into a knowledge graph; \textit{Artifact Design and Development} builds four automated pipelines that extract and enrich data from repositories, libraries, and community platforms, forming a domain-specific knowledge graph for inference and recommendation; \textit{Demonstration} applies the framework, via its prototype implementation, \textsc{ModelSelect}, to real-word applications using a curated dataset of representative case studies; \textit{Evaluation} assesses pipeline accuracy, validates the knowledge graph, and compares \textsc{ModelSelect} with Generative AI tools using metrics such as coverage and overlap; and \textit{Communication} documents and shares the research design, methodology, and findings with academic and practitioner communities, through publicly available datasets and evaluation materials.


\subsubsection{Literature Study to Support Design Decisions}

A semi-structured literature review was conducted using the snowballing methodology proposed by \cite{wohlin2014snowballing}. The review examined existing AI model selection practices and MCDM frameworks. Insights from this study guided the design of key components, including the metadata schema, feature extraction strategies, and the knowledge graph structure. Detailed findings are detailed in Section \ref{related_work}, with additional procedural information in our Mendeley Data Repository \cite{modelselectmendeley}.

\subsubsection{Evaluation of Data Collection Pipelines}
The automated data collection pipelines were evaluated through experiments to measure their accuracy and recall. Six key tasks were validated: dependency extraction, which identifies and resolves imported libraries in AI repositories; AI library identification, which identifies and labels AI libraries that implement models; Library models, which extracts models from library documentation and metadata; Base models, which maps model variations to their base models; sentiment analysis, which analyzes developer community reviews to detect positive/negative feedback; and Quality extraction, which detects software qualities based on ISO/IEC-25010 \cite{ISO_IEC_25010_2023} in developer reviews.

\smallskip

These experiments ensured that the pipelines could reliably collect and normalize metadata from diverse sources, as outlined in Section \ref{eval:pipeline_manual_eval}.

\subsubsection{Evaluation of Data Integration, Inference Engine, and Practical Utility}
The quality and practical utility of \textsc{ModelSelect} were evaluated through comparative analysis using a dataset of real-world case studies. \textsc{ModelSelect} recommendations were compared against those produced by four GenAIs, including GPT-4o \cite{openai2024gpt4ocard}, GPT-5 mini \cite{openai2025gpt5system}, Gemini 2.5 Flash \cite{comanici2025gemini25pushingfrontier}, and Claude Sonnet 4.5 \cite{anthropic2025claudesonnet45}, with the evaluation focusing on coverage and recommendation overlap percent.

\subsection{Case Study Methodology}

The framework evaluation followed Yin’s case study methodology \cite{yin_case_study_guidelines}, which emphasizes systematic procedures for empirical research. This approach included defining clear objectives, using multiple sources of evidence, and following a structured protocol to ensure reliability and validity. To validate the framework, we constructed a curated dataset of case studies drawn from high-quality peer-reviewed publications. The selected studies and their associated code repositories were closely examined for the AI models and libraries used, their applications, and the stated rationales for their selection. This information was compiled into an annotated table that maps rationales to specific models. This evidence-based dataset serves as a reference for evaluating the framework. The paper selection and curation processes are elaborated in Section \ref{subsection:casestudy_dataset}.

\section{ModelSelect Framework}
\label{modelselect_framework}


Our proposed design is comprised of four interconnected pipelines, with each addressing a specific aspect of the AI model selection process: 
\begin{enumerate*}[label=\arabic*)]
    \item GitHub Repository Data Extraction,
    \item AI Library Data Extraction,
    \item Model-Feature Identification,
    \item Quality Assessment Data Extraction.
\end{enumerate*}

\medskip

\begin{figure}[t!]
    \centering
    \vspace{-2mm} 
    \includegraphics[width=\columnwidth,keepaspectratio]{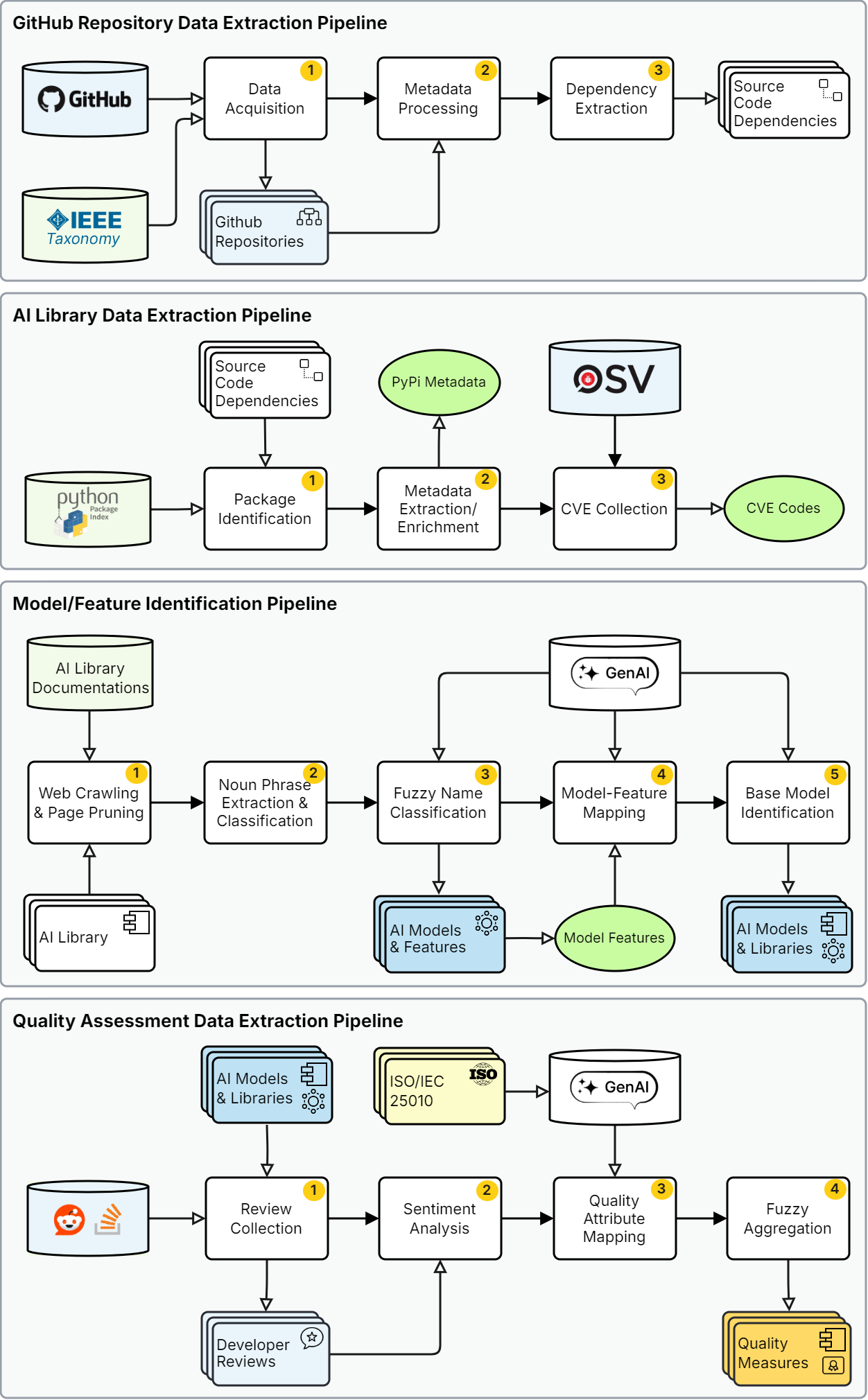}
    \caption{\textsc{ModelSelect} Pipelines for data acquisition, enrichment, and mapping}
    \label{fig:pipeline}
    \vspace{-2mm} 
\end{figure}

Figure \ref{fig:pipeline} outlines the design of our framework and summarizes the activities for each pipeline, highlighting the inputs, and outputs for each activity. Together, these four pipelines form the core operational workflow of the \textsc{ModelSelect} framework. They transform heterogeneous data sources into a unified, queryable knowledge base for AI model selection. The aggregated outputs are integrated into the inference model, enabling efficient retrieval and reasoning over the collected data.

\smallskip

To ensure data integrity and verify the accuracy of the automated extraction processes, an evaluation phase was conducted. A representative sample (10-20\%) of each pipeline's outputs underwent manual review. During this process, the pipeline tasks were manually replicated, and the curated data were compared against the automated outputs to quantify precision and identify discrepancies. The following sections detail the pipelines, the inference model architecture, and the methodology used for this evaluation.

\subsection{\textsc{ModelSelect} Pipelines}
\label{subsection:pipelines}

\subsubsection{GitHub Repository Data Extraction}

This pipeline involves extracting data from GitHub repositories related to AI. Initially, repositories are collected along with key metadata, including repository name, URL, description, number of stars, forks, size, programming language, contributor count, creation date, and last update date. These repositories are then categorized using IEEE taxonomy terms \cite{IEEEThesaurus, IEEEtaxonomy}. To ensure high relevance and quality, the metadata is processed and filtered using median thresholds and multiple criteria. This step removes unpopular or obscure repositories. Next, the codebases of the remaining repositories are analyzed to extract their dependencies by identifying imported libraries.

\smallskip

The output of this pipeline is a curated dataset containing detailed information about AI-related repositories and their dependencies. This dataset serves as the foundation for subsequent pipelines.


\subsubsection{AI Library Data Extraction}
The AI library data extraction pipeline aims to analyze further the dataset obtained from the previous pipeline. It analyzes source code dependencies to identify Python AI libraries associated with them. Once identified, metadata for each library is retrieved from PyPI (Python Package Index). This includes details such as library name, homepage, description, tags, and current version. Additionally, security vulnerabilities (CVEs) associated with each library are identified using the OSV API.

\smallskip

The output of this pipeline is a dataset of shared libraries imported by the AI repositories, enriched with metadata and security details.

\subsubsection{Model/Feature Identification}
The model/feature identification pipeline processes libraries from the previous pipeline to extract the AI models they implement or support. A subset of libraries from the prior pipeline that are tagged or denoted as AI-related is selected for further analysis. Additional documents, such as PyPI descriptions, homepages, and websites, are collected for these libraries. These documents are then cleaned, categorized, and structured for processing.

\smallskip

Key noun phrases from the textual documents that may refer to AI models or features are extracted. Each noun phrase is labeled as model, feature, or neither using advanced GenAI capabilities and fuzzy classification. GenAI tools are also used to extract definitions and map models to their features based on co-occurrences of the noun phrases. Model variations are clustered into categories by identifying base model names and related features.

\smallskip

The pipeline produces a two-way index that maps AI libraries to AI models and vice versa. This index includes definitions of base models and their variations, offering detailed information on the capabilities and features of AI libraries for implementing specific models.

\subsubsection{Quality Assessment Data Extraction}
The quality assessment data extraction pipeline enriches the collected dataset by assessing the quality attributes of each AI model-library pair. While the previous pipelines provide metadata and technical details, RSEs may struggle to decide because they lack the expertise to compare them effectively. This pipeline addresses that challenge by assessing software quality attributes to provide additional insights.

\smallskip

To achieve this, online software forums such as Stack Overflow and Reddit are searched for reviews of implementations of AI model variations using specific libraries. GenAIs are then used to map positively and negatively charged sentences in these reviews to software quality attributes based on ISO/IEC 25010 \cite{ISO_IEC_25010_2023} definitions. Standard definitions of quality attributes are provided as context to the GenAI to ensure accurate mapping. The pipeline utilizes fuzzy aggregation methods to calculate overall sentiment scores and review counts for each quality attribute. 

\smallskip

The results are appended to the indexes generated in the model/feature identification pipeline. This provides researchers with a comprehensive view of the strengths and weaknesses of each AI model-library pair, supporting informed decision-making.

\subsection{Knowledge Base}
\label{subsection:knowleedge_base}
\begin{figure}[t!]
    \centering
    \includegraphics[width=\linewidth]{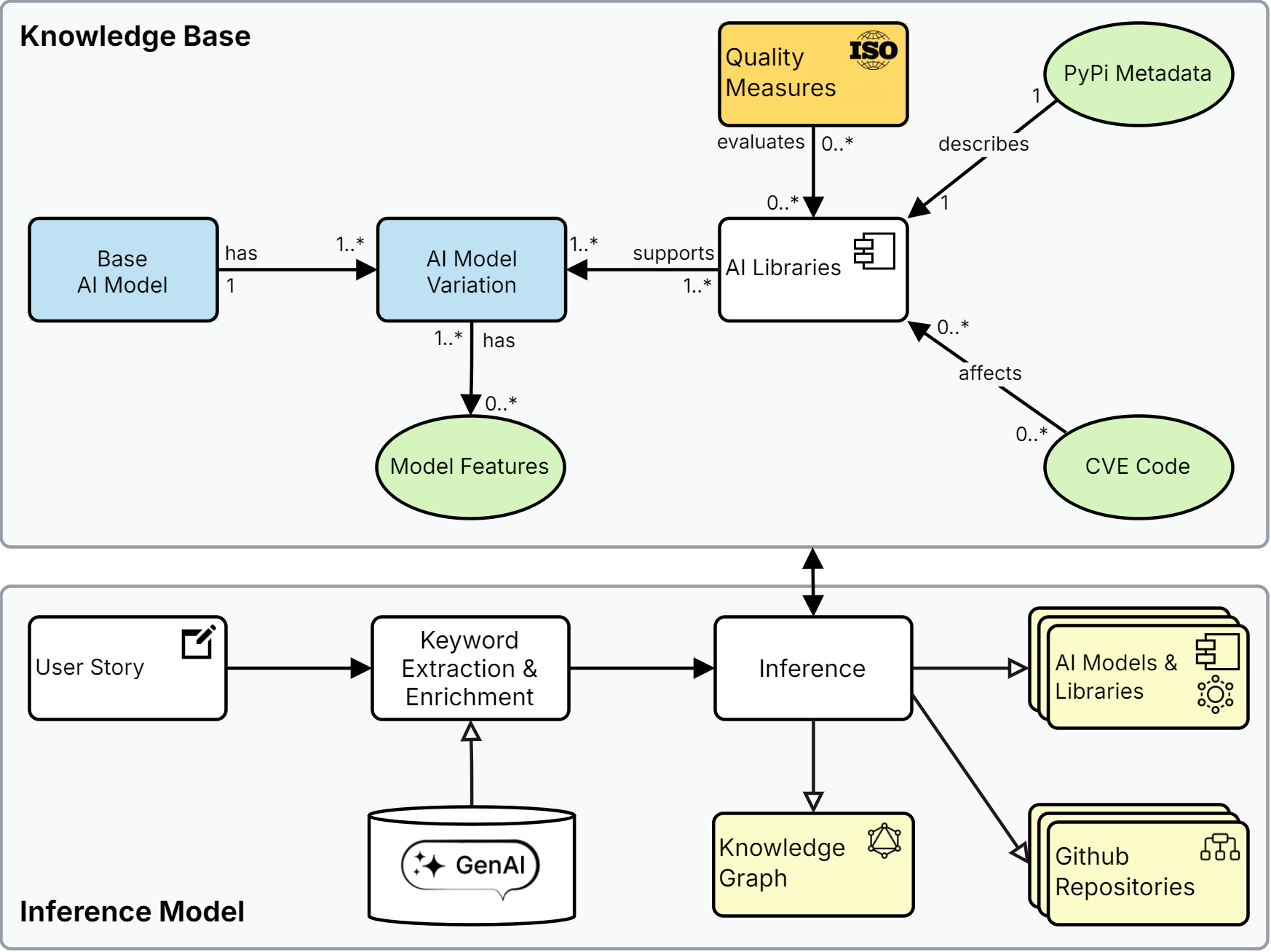}
    \vspace{-4mm}
    \caption{\textsc{ModelSelect}'s knowledge base and inference model.}
    \label{fig:knowledgebase}
\end{figure}

The knowledge base organizes interconnected entities and facilitates comprehensive analysis of AI models and libraries. Thus, it serves as the structural foundation of the \textsc{ModelSelect} system. This knowledge base is stored as a set of optimized Elasticsearch \cite{noauthororeditor2015elasticsearchelasticsearch} indices. As illustrated in Figure \ref{fig:knowledgebase}, the schema links \textit{Base AI Models} to their corresponding \textit{AI Model Variations}, This allows different implementations, architectures, or configurations under a unified conceptual model. Each variation may contain a unique set of \textit{Features} that describe its capabilities, usage constraints, or architectural characteristics.

\smallskip

\textit{AI Libraries} are associated with model variations they support, creating a direct mapping between conceptual models and concrete implementations. Libraries are enriched with metadata from external sources such as PyPI and GitHub repositories that import them, including keywords and descriptions, popularity statistics (i.e., stars and forks), and dependency links. Additional entities include \textit{Quality Measures} and \textit{CVE Codes}, which may impact a library's use in production systems.

\smallskip

\begin{figure}[!ht]
    \centering
    \includegraphics[width=1\linewidth]{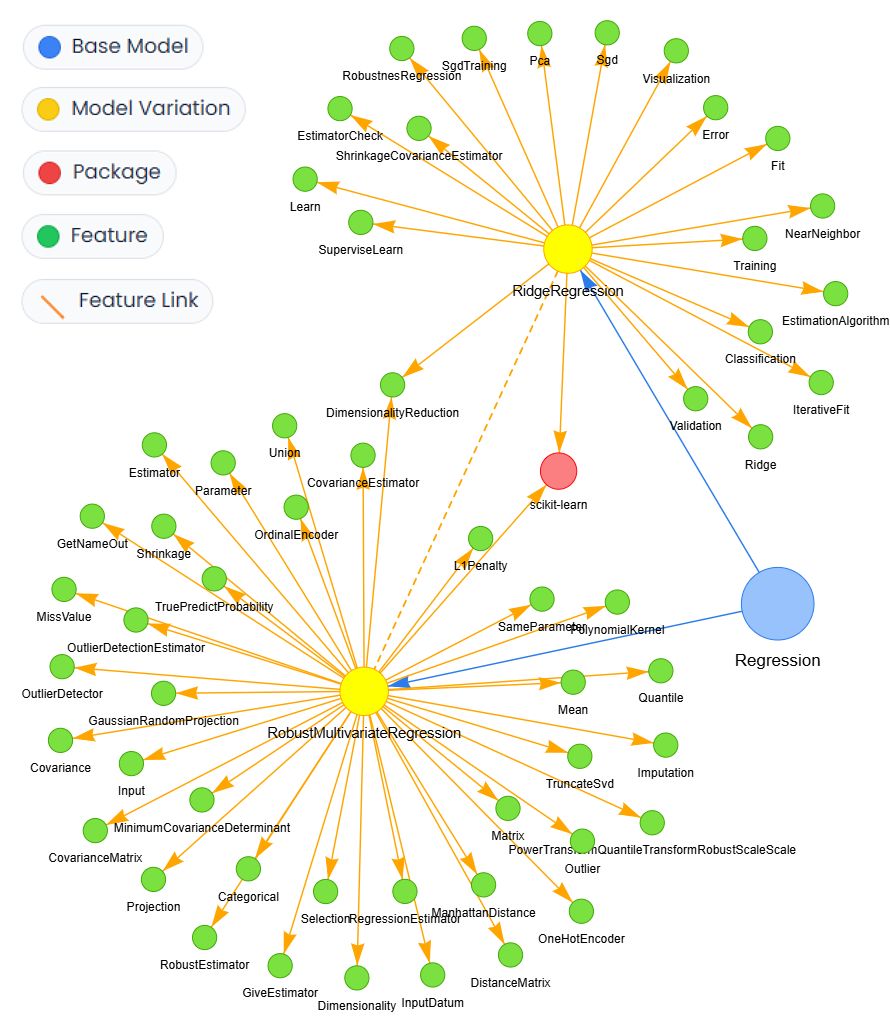}
    \caption{Example representation of a knowledge graph comparing two Regression model variations along with their features}
    \label{fig:knowledgebasegraph}
\end{figure}

Figure \ref{fig:knowledgebasegraph} illustrates a subgraph from the knowledge base, showing relationships between a base model, a subset of its variations, and their features. In this example, the base model, \textit{Regression}, is represented in blue, while two chosen variations, \textit{Ridge Regression} and \textit{Robust Multivariate Regression}, are depicted in yellow. Each variation is linked to its respective features, shown in green, which describe the model's functionalities and attributes. Additionally, the supporting library \texttt{scikit-learn} is highlighted in red, indicating its support for both variations.

\smallskip

This interconnected knowledge base captures relationships between models, their implementations, supporting libraries, vulnerabilities, and performance evaluations. By structuring information in this way, the knowledge base enables precise, explainable inference for decision-making and advanced comparisons.

\subsection{Inference Model}
The inference model integrates all collected data into an Elasticsearch \cite{noauthororeditor2015elasticsearchelasticsearch} database for efficient indexing and retrieval. It connects information about repositories, libraries, AI models, and user reviews, creating rich connections between these entities.

A specialized query engine is implemented to process detailed paragraphs about the user's objectives and requirements. The engine first extracts relevant keywords, removes trivial or overly domain-specific terms, and enriches the input with semantically related concepts. This refined set of keywords drives a multi-field Elasticsearch query, which matches model names, definitions, known variations, and library attributes. Ultimately, the inference model returns a ranked list of models and their variations, along with the libraries that implement them, the features of each pair, and the evidence supporting them.

The interconnected dataset provides highly specific and informative results, which can be visualized in various formats to support informed decision-making. This engine ensures that the entire system is accessible, efficient, and capable of addressing complex user queries. As mentioned in Section \ref{introduction:prototype}, a recommender system representing an implementation of this inference model is publicly available, which enables users to explore AI model recommendations through a rich and evidence-driven interface.

\section{Metadata Analysis}
\label{metadata}

\subsection{Pipeline Results}

This section discusses pipeline results, providing a quantitative overview of the scale and coverage achieved at each processing stage.

\smallskip
In the GitHub Repository Extraction pipeline, 37,176 repositories were crawled, of which 4,610 were selected based on predefined inclusion criteria. Within these, 8,421 dependencies were extracted, corresponding to 2,001 distinct PyPI libraries.

\smallskip
The AI Library Data Extraction pipeline identified 663 AI-tagged PyPI libraries and 109 additional libraries, which were classified as AI-related through additional heuristics based on their summaries and homepage texts. From these libraries, the crawler processed 15,113 documentation URLs, averaging 19.58 per library, indicating relatively rich documentation coverage.

\smallskip
In the Model-Feature Extraction pipeline, 250 libraries were identified as being associated with at least one AI model, comprising 89 distinct base AI models and 1,852 variations. This step required substantial LLM-driven processing, resulting in 133,880 LLM requests to identify and map models and features.

\smallskip
Quality Assessment Extraction involved the processing of 55,149 reviews, from which 7,549 quality scores were successfully computed. This phase, also, was computationally intensive, generating 63,414 LLM requests.

\smallskip
Overall, these metrics indicate that the pipeline is capable of operating at a large scale while maintaining detailed coverage of AI model-library relationships. The relatively high counts of processed URLs, reviews, and LLM requests highlight both the density of the collected data and the effectiveness of GenAIs in processing heterogeneous textual sources.

\subsection{Prevalence of Supported AI Models}
The analysis of the top 10 most supported AI models reveals a clear dominance of \textit{Transformer}-based architectures, with 103 libraries in our dataset supporting them. This is followed by the \textit{Miscellaneous} category (89) and \textit{Convolutional Neural Networks (CNNs)} (76), both of which are widely implemented across various libraries, highlighting their versatility and continued relevance in diverse AI applications. Traditional techniques such as \textit{Boosting} (28), \textit{Support Vector Machines (SVMs)} (28), and \textit{Language Models} (26) show more moderate representation, indicating that despite the advancements in Transformer architectures, more conventional model families remain essential in many domains. In contrast, approaches such as \textit{Autoencoders} (24), general \textit{Deep Learning} methods (23), classical \textit{Neural Networks (NNs)} (22), and \textit{Decision Trees} (20) are supported by fewer libraries, suggesting either more specialized use cases or reduced emphasis in contemporary AI library development. Overall, the distribution underscores the strong industry and research momentum behind Transformer architectures while also reflecting the persistence of foundational ML techniques in the AI ecosystem.

\subsection{AI Model Variations}
The analysis of the base models and their variations shows that, while \textit{Transformer} architectures (304 variations) dominate current AI trends, the \textit{Miscellaneous} category has the highest variation count (335), suggesting a wide diversity of less standardized or hybrid model types in active use across AI Libraries. \textit{CNNs} remain a strong presence, with 294 variants, reflecting their ongoing relevance for image-based and spatial data tasks. \textit{Language Models} (89) and general \textit{Deep Learning} frameworks (83) also maintain notable representation, while traditional approaches such as \textit{Boosting} (40), \textit{Regression} (35), \textit{Autoencoders} (34), \textit{Neural Networks (NNs)} (26), and \textit{Generative Adversarial Networks (GANs)} (25) still occupy meaningful positions within the ecosystem, despite their more modest variation counts, indicating the coexistence of both modern and foundational paradigms in AI development

\subsection{Top AI Libraries}
The analysis of AI Libraries supporting the broadest range of models reveals a highly skewed distribution, with a few dominant platforms accounting for the majority of coverage. \texttt{huggingface\_hub} emerges as the most comprehensive AI library, supporting the most significant number of models (512) and significantly surpassing all other libraries in our dataset. \texttt{torch} is the second-most widely supported library, with 326 models. Mid-tier libraries, such as \texttt{spacy} (156), \texttt{scikit-learn} (125), and \texttt{tensorflow} (90), demonstrate moderate yet substantial coverage, indicating a balance between specialized functionality and broad applicability. At the lower end, other libraries support fewer than 80 models, typically focusing on niche domains or specialized hardware integration. This distribution underscores the centralization of AI model support around a small number of widely adopted and actively maintained libraries. At the same time, a long tail of more specialized AI Libraries serves targeted use cases.

\section{Evaluation}
\label{evaluation}

\begin{table*}[t!]
\centering
\scriptsize
\renewcommand{\arraystretch}{1.3}
\label{table:pipelineExperiments}
\begin{tabularx}{\linewidth}{|l|l|Y|Y|Y|Y||Y|Y|Y|Y||Y|Y|Y|Y|}

\cline{1-14}

\multirow{2}{*}[-8ex]{\textbf{Pipeline}} & \multirow{2}{*}[-8ex]{\textbf{Experiment}} & \multicolumn{4}{c||}{\textbf{Precision (\%)}} & \multicolumn{4}{c||}{\textbf{Recall (\%)}} & \multicolumn{4}{c|}{\textbf{F-1 Score (\%) }} \\ \cline{3-14}

& & \rotatebox[origin=c]{90}{ \parbox{2.2cm}{\centering Gemini 2.5 Flash}} & \rotatebox[origin=c]{90}{ \parbox{2.2cm}{\centering GPT-5 mini }} & \rotatebox[origin=c]{90}{ GenAI Fused } & \rotatebox[origin=c]{90}{ \parbox{2.2cm}{\centering \textsc{ModelSelect}}} 
& 
\rotatebox[origin=c]{90}{ \parbox{2.2cm}{\centering Gemini 2.5 Flash }} & \rotatebox[origin=c]{90}{ \parbox{2.2cm}{\centering GPT-5 mini }} & \rotatebox[origin=c]{90}{ GenAI Fused } & \rotatebox[origin=c]{90}{ \parbox{2.2cm}{\centering \textsc{ModelSelect} }}
& 
\rotatebox[origin=c]{90}{ \parbox{2.2cm}{\centering Gemini 2.5 Flash }} & \rotatebox[origin=c]{90}{ \parbox{2.2cm}{\centering GPT-5 mini }} & \rotatebox[origin=c]{90}{ GenAI Fused } & \rotatebox[origin=c]{90}{ \parbox{2.2cm}{\centering \textsc{ModelSelect} }} \\ \cline{1-14}

GitHub Repository Extraction & Dependency Extraction & 18 & 28 & 17 & \textbf{82} 
                                   & 27 & 23 & 34 & \textbf{96} 
                                   & 22 & 25 & 23 & \textbf{88} \\ \cline{1-14}

AI Library Data Extraction & AI Library Identification & 79 & 79 & 79 & \textbf{86} 
                                       & 95 & 96 & 96 & \textbf{84} 
                                       & 92 & 94 & 93 & \textbf{93} \\ \cline{1-14}

\multirow{2}{*}{Model-Feature Extraction} & Library Models & 88 & 91 & 88 & \textbf{81}
                                             & 62 & 44 & 64 & \textbf{70}
                                             & 62 & 50 & 65 & \textbf{75} \\ \cline{2-14}
                            & Base Models & 96 & 94 & 95 & \textbf{79}
                                          & 74 & 77 & 82 & \textbf{80}
                                          & 84 & 85 & 88 & \textbf{79} \\ \cline{1-14}

\multirow{3}{*}{Quality Assessment Extraction} & Postive Sentiments & 98 & 96 & 94 & \textbf{87}
                                                 & 45 & 46 & 51 & \textbf{82}
                                                 & 61 & 62 & 66 & \textbf{89} \\ \cline{2-14}
                            & Negative Sentiments & 87 & 94 & 95 & \textbf{87}
                                                  & 82 & 55 & 82 & \textbf{89}
                                                  & 84 & 70 & 84 & \textbf{88} \\ \cline{2-14}
                            & Qualities & 33 & 38 & 40 & \textbf{77}
                                        & 35 & 40 & 38 & \textbf{70}
                                        & 27 & 35 & 32 & \textbf{71} \\ \cline{1-14}

\end{tabularx}
\caption{Pipeline Experiments Results}
\end{table*}

\subsection{Pipeline Evaluation}
\label{eval:pipeline_manual_eval}

The performance of each pipeline was evaluated using a human-annotated gold-standard approach. For each pipeline, two independent evaluators reviewed a random sample of 10\% of the total data rows processed during execution. The evaluators manually performed the tasks assigned to the pipeline. These human-generated outputs were consolidated into a ground-truth dataset for performance analysis and comparison. Rows in the ground truth dataset were then compared with outputs of \textsc{ModelSelect} (using LLaMa-3.2:70b \cite{Llama3Herd}) and two GenAIs (Gemini 2.5 Flash \cite{comanici2025gemini25pushingfrontier} and GPT-5 mini \cite{openai2025gpt5system}). The effect of combining the results of both GenAIs (GenAI Fused) has also been considered to assess their collective performance.

\smallskip

The evaluation results for all pipelines and their experiment are summarized in Table \ref{table:pipelineExperiments}. Most annotated entries were consistent, indicating that predictions were closely aligned in each category. Classification metrics —precision, recall, and F1-score—were calculated for all experiments. Overall, the results show consistent performance and low variance in precision and recall across different experiments. They also suggest that the procedures embedded in different pipelines can serve their purpose well, with only a small proportion of misclassifications observed in edge cases, which will be discussed in Section \ref{insights}.

\medskip
\subsubsection{GitHub Repository Data Extraction}

In this pipeline, the automated system identifies Python dependencies in AI-related repositories using optimized heuristics to quickly detect imported libraries without exhaustively parsing all source files. To evaluate this process, two researchers independently performed the same dependency identification task on the sampled repositories with the manual method of their choice. The pipeline outputs were then compared with the human-generated lists to estimate the pipeline's accuracy. Results show that \textsc{ModelSelect} is capable of thoroughly crawling AI repositories and extracting dependencies with high precision (82\%) and recall (96\%). GenAIs, however, yield unsatisfactory results in this task, with low precision and recall values (28\% and 34\%, respectively). 

\medskip
\subsubsection{AI Library Data Extraction}

In the AI Library Data Extraction pipeline, Python dependencies originating from the previous pipeline are examined to identify corresponding entries in PyPI and extract their metadata. All libraries are tagged as AI-related or not using a binary variable. Two researchers independently classified a random subset of these libraries, and their classifications served as the ground truth for calculating the pipeline's precision, recall, and F1-score. Results indicate the effectiveness of the pipeline's heuristic method for this task, which is comparable to that of GenAIs. 

\medskip
\subsubsection{Model-Feature Identification}

The Model-Feature Extraction pipeline is the most essential part of the framework, which extracts supported AI models and their features from the AI library documentation. The evaluation leveraged the fact that each automated extraction can be traced to the specific URLs and their text fragments. Human evaluators were provided with these source documents, and they extracted the model names. They were tasked with labeling each extracted instance as \textit{Accurate} or \textit{Not Accurate}. This review process enabled a focused assessment of the accuracy of noun phrase extraction, semantic classification, and the correct identification of model variations. According to the metrics, 80\% (precision) of library-model associations done by \textsc{ModelSelect} are correct. GenAIs also exhibit similar performance, with 88\% and 91\% precision.

\smallskip

To ensure that this pipeline's categorization of variations into base models meets standards, another experiment was conducted. In this experiment, base-variation model pairs were evaluated by human evaluators. While \textsc{ModelSelect}'s precision (79\%) and recall (80\%) show moderately good placement of model variations under base model categories, there is room for improvement in this regard.

\medskip
\subsubsection{Quality Assessment Data Extraction}

A similar validation approach was applied to the Quality Assessment Extraction pipeline, which infers software quality attributes by analyzing sentiment-bearing sentences from online user reviews of AI models and libraries. The sampled results (including extracted sentences, assigned sentiment polarity, and mapped quality attributes) were presented to human evaluators for verification. This allowed the assessment of both the sentiment classification accuracy and the correctness of quality mapping against qualities mentioned in ISO/IEC 25010 \cite{ISO_IEC_25010_2023}. While GenAIs excelled at precisely classifying sentiments, their recall for sentences with positive sentiment (51\% at most) and overall performance in extracting qualities were low.

\subsection{Case-Study Dataset}
\label{subsection:casestudy_dataset}

\begin{table}[t!]
\centering
\scriptsize
\renewcommand{\arraystretch}{1.4}
\label{table:evDomainCount}
\begin{tabularx}{\linewidth}{|l|Y|Y|}
\hline
\textbf{Domain} & \textbf{\# of Case Studies} & \textbf{\# of Models} \\ \hline
Biology & 5 & 9\\ \hline
Computer Science and Security & 8 & 15 \\ \hline
Ecology \& Environmental Science & 5 & 5 \\ \hline
Environmental Modeling \& Software & 8 & 11 \\ \hline
Healthcare & 15 & 18 \\ \hline
Physical Sciences & 5 & 5 \\ \hline
Vision and Robotics & 4 & 4 \\ \hline
\textbf{Grand Total} & \textbf{50} & \textbf{40} \\ \hline
\end{tabularx}
\caption{Distribution of Case Studies Across Domains}
\end{table}

To construct a robust evaluation dataset, we curated over 200 peer-reviewed works from well-established venues relevant to RSE and AI applications. Papers were identified using systematic keyword searches in databases such as ScienceDirect, IEEE Xplore, and ACM Digital Library, with filters applied to focus on publications from 2021 onward to ensure relevance to current practices. These works were required to involve AI models applied to domain-specific problems, to provide sufficient metadata (e.g., title, authors, DOI, repository links), and to originate from venues ranked Q1 or Q2 or with high citation impact. Papers were excluded if they lacked relevance to RSE or had no available code repositories.

\smallskip

From this initial pool, a multi-stage selection process was applied to refine the dataset, ensuring diversity across application domains such as healthcare, biology, and physical sciences. Table \ref{table:evDomainCount} describes the number of case-studies belonging to each domain. During the selection process, each work was manually reviewed to extract information on the AI models used, the rationale for their selection (e.g., application context, paradigm, and data type), and the libraries employed for implementation. Works that lacked sufficient explanation of model choice or implementation details were also excluded.

\smallskip

These efforts yielded a final dataset of real-world case studies, comprising 50 studies and over 90 documented instances of AI model usage. Each instance includes critical metadata, such as the model name, application context, selection rationale, and supporting evidence. For all items, code repositories and scripts were manually examined to identify the libraries used, and this information was later used to evaluate library recommendations for specific models.

More details about the dataset composition, definitions, and process are presented in the Mendeley Data repository \cite{modelselectmendeley}, which further describes the selection procedure and inclusion/exclusion criteria.

\subsection{Baselines}

 To evaluate, compare, and contextualize the findings, four state-of-the-art AI agents (namely, GPT-4o \cite{openai2024gpt4ocard}, GPT-5 mini, Gemini-2.5-Flash \cite{comanici2025gemini25pushingfrontier}, and Claude Sonnet 4.5 \cite{anthropic2025claudesonnet45}) were queried to suggest AI models and libraries for each of the 94 test cases using the rationale for selection as the query input. Their recommendations were cross-referenced both with the literature-based benchmark (to evaluate their alignment with human experts), and with \textsc{ModelSelect}’s recommendations (to measure the degree of overlap with our method’s outputs). This dual comparison enabled a comprehensive quantitative assessment of how closely the AI agents reflected both empirical evidence and our systematic approach.

\subsection{Evaluation Metrics}

To assess the effectiveness of our proposed design, we made use of two complementary metrics: 

\begin{enumerate}
    \item \textbf{\textsl{Coverage:}} evaluates whether the correct choice appears among the top-$k$ recommendations (with $k=10$ being specifically utilized). This metric can be applied both to model and library recommendations.
    
    \smallskip
    \item \textbf{\textit{Overlap:}} quantifies what percentage of the models suggested by each GenAI for a given rationale also existed in recommendations of \textsc{ModelSelect} for the same rationale. 
\end{enumerate}

\subsection{Results}

\begin{table}[t!]
    \centering
    \scriptsize
    \renewcommand{\arraystretch}{1.2}
    \label{table:results}
    \begin{tabular}{|c|c|c|c|c|c|c|c|}
        \hline
        \multirow{2}{*}{\vspace{-8em} \textbf{Context}} & \multirow{2}{*}{\vspace{-8em} \textbf{Metric}} & \multicolumn{5}{c|}{\textbf{Method}} \\ \cline{3-7}
        & & \rotatebox[origin=c]{90}{ \parbox{2.2cm}{\centering  GPT-4o}} & \rotatebox[origin=c]{90}{ \parbox{2.2cm}{\centering GPT-5 mini}} & \rotatebox[origin=c]{90}{ \parbox{2.2cm}{\centering Gemini 2.5 Flash}} & \rotatebox[origin=c]{90}{ \parbox{2.5cm}{\centering  Claude Sonnet 4.5}} & \rotatebox[origin=c]{90}{ \parbox{2.5cm}{\centering \textsc{ModelSelect}}} \\ \hline

        \multirow{2}{*}{Models} & Coverage (\%) & 82.61 & 68.48 & 77.17 & 89.13 & 86.96 \\ \cline{2-7}
        & Overlap (\%) & 53.48 & 54.57 & 47.83 & 51.59 & - \\ \hline
        Libraries & Coverage (\%) & 85.87 & 86.96 & 90.22 & 91.30 & 82.61 \\ \hline

    \end{tabular}
    \caption{Case-Study Evaluation Results on Model and Library Recommendations}
\end{table}

The evaluation results, summarized in Table \ref{table:results}, highlight the effectiveness of our proposed approach in delivering accurate and transparent model and package recommendations across diverse rationale categories. Additionally, a detailed breakdown of results regarding model recommendations for each rationale category is provided in \cite{modelselectmendeley}.

\smallskip

Notably, in the model recommendation task, our method achieves reasonable coverage, with an average score of 86.96\%, comparable to that of GenAIs. While the results show strong performance in identifying accurate model recommendations, the proposed framework emphasizes not only accuracy but also the ability to provide traceable, explainable recommendations.

\smallskip

The overlap metric, on the other hand, measures how well the recommendations from each GenAI align with those of \textsc{ModelSelect}. By measuring overlap, we can assess the extent to which GenAIs independently converge on the same recommendations, providing indirect validation of the robustness of our design. The results in Table \ref{table:results} show that overlap scores vary across systems, with GPT-5-mini having the highest average overlap at 54.57\%, followed by GPT-4o at 53.48\%. However, the overlap metric also identifies areas where divergence occurs, reflecting the differences in how generative AI systems infer models compared to our approach.

\smallskip

At a glance, results for the library recommendation task show that all evaluated methods are effective at recalling suitable libraries for each model, with coverage scores exceeding 80\% across the board. Claude Sonnet 4.5 achieved the highest coverage (91.30\%), followed closely by Gemini-2.5 Flash (90.22\%), GPT-5 mini (86.96\%), and GPT-4o (85.87\%). \textsc{ModelSelect}, while at 82.61\%, remains consistent. The modest gap can be attributed to our data collection strategy, which records only explicit evidence of a model being implemented with a specific library. This results in a more conservative, evidence-based measure.






\section{Discussion}
\label{discussion}


\subsection{Research Questions Revisited}

The research questions were formulated to address the challenges outlined in problem formulation (Section \ref{problem_formulation}), namely the lack of scalable, reproducible, and context-aware approaches to AI model selection for research software engineering tasks. Each question aligns with a key stage in the design, instantiation, and evaluation of the proposed \textsc{ModelSelect} framework. Specifically, \hyperref[RQ1]{RQ1} examines the feasibility of automating the collection of relevant model metadata and usage rationales. \hyperref[RQ2]{RQ2} investigates how such data can be structured into a formalized knowledge representation that supports evidence-based recommendations. Finally, \hyperref[RQ3]{RQ3} evaluates the effectiveness of the resulting decision model in producing accurate, domain-aware, and rationale-justified recommendations. Collectively, these questions serve to validate the framework's effectiveness, transparency, and practical value in assisting RSEs with model selection.


\subsubsection{RQ1} This question was addressed through the design and implementation of an automated pipeline that gathers AI model metadata and features from multiple heterogeneous sources, including AI repositories, official library documentation, and community platforms (Section \ref{subsection:pipelines}). The pipeline extracts and enriches the desired information in a transparent and traceable manner. Results of evaluation (Section \ref{eval:pipeline_manual_eval}) against a gold-standard annotation set yielded high inter-rater reliability, indicating that the automated process retained accuracy while scaling well beyond manual collection.


\subsubsection{RQ2} To answer this question, the collected metadata was modeled as a knowledge graph (Section \ref{subsection:knowleedge_base}) linking AI base models, their variations, associated libraries, features, and extracted rationales. This structure supports many-to-many relationships between models and their features, enabling traceability and transparency in the recommendation process. An inference mechanism, informed by MCDM principles, traverses this graph to identify candidate models aligned with a user's stated functional and non-functional requirements, while preserving the provenance of each recommendation.


\subsubsection{RQ3} The effectiveness of \textsc{ModelSelect} was assessed through controlled experiments on an evaluation case-study dataset annotated with model and library choices (Section \ref{evaluation}). The model was evaluated based on its Top-10 ranking performance and the rationale alignment with human assessments. Comparative analysis against four GenAIs demonstrated that the proposed method outperformed them in delivering domain-aware, context-specific, and rationale-supported recommendations. The results show that integrating structured metadata with empirical reasoning yields more accurate, explainable model selection than relying on general-purpose GenAIs.


In summary, the systematic exploration of these research questions confirms that automated metadata extraction, structured knowledge representation, and an MCDM-based inference process can collectively provide reproducible, transparent, and practical support for AI model selection in research software engineering contexts.

\subsection{Insights}
\label{insights}

\subsubsection{Advantages of Structured Systems}

Compared to free-form generative responses, structured systems produce more consistent, explainable outputs \cite{he2024template}. This is evident in the results of the "Dependency Extraction" experiment in the GitHub repository data extraction pipeline, where GenAIs' performance values were considerably lower. Structured systems use strict decision boundaries and can minimize divergence between multiple AI runs, thereby improving reliability and alignment with human evaluators.

\subsubsection{Sensitivity of GenAIs to Prompting}

 Revisiting the results for pipeline experiments, recall values for some experiments were considerably lower than those for others. This indicates that large language models are susceptible to prompt phrasing and task description \cite{sun2024evaluating, ceballos2024open}, such that even minor variations in template \cite{he2024template}, format or punctuation marks \cite{seleznyov2025punc} in prompts can lead to divergent interpretations of requirements. In some cases, such as the "Library Models" and "Qualities" experiments, this led to inconsistent entity extraction or classification labels, reflecting the inherently probabilistic nature of generative inference. Such variability highlights the need for controlled prompt engineering to ensure reproducible outputs. To this end, in this work, each prompt was asked three times from GenAIs, with the majority of responses being chosen as their final answer.

\subsubsection{Occurrence of Hallucinated Model Names}

In specific experiments, GenAI outputs contained references to model names or AI libraries not present in the dataset. This led to a lower-than-normal performance in model/feature identification pipeline experiments. Although infrequent, these hallucinations suggest that generative models occasionally rely on prior knowledge patterns rather than the provided context, especially when uncertain.

\subsubsection{Data Scope and Completeness}
The completeness of the collected metadata inherently limits the size of the model pool. Our collection process relies heavily on open-source sources, and therefore, undocumented models or features may be absent from the knowledge base. While the framework can identify highly customized model variations and the millions of variations in model repositories, this work does not delve deeply into the hierarchical structure of models. This is a known trade-off and should be interpreted in the context of this work's primary objective: introducing an accurate and extensible framework, rather than releasing a fully exhaustive tool. Nonetheless, these limitations highlight the shortcomings of \textsc{ModelSelect} in certain rationales, as presented in the evaluation results in \cite{modelselectmendeley}.

\subsubsection{Future Extensions}
Importantly, the proposed framework is designed to be re-executed regularly, allowing for continual enrichment of the knowledge base as new sources and documentation become available. Future iterations could integrate curated or proprietary datasets, along with mechanisms for user-driven contributions, to further expand completeness and diversity.

\subsection{Limitations and Threats to Validity}

\subsubsection{Construct Validity}
Several critical tasks, including model/feature identification/mapping, definition extraction, and sentiment-to-quality mapping, utilize GenAIs. While our experiments demonstrated strong reliability on both final pipeline outputs and individual LLM tasks, and while manual verification of approximately 10-20\% of the results was conducted, dependence on LLMs introduces a black-box element into the process. These models may hallucinate, misclassify entities, or inherit latent biases from their training data, which can in turn propagate into downstream recommendations.

\smallskip

Furthermore, the use of community sentiment to infer ISO/IEC 25010 \cite{ISO_IEC_25010_2023} quality attributes relies on the assumption that informal user feedback correlates with formally defined software quality dimensions. As prior work suggests, community discussions often conflate subjective usability impressions with objective technical performance, introducing potential measurement noise and construct misalignment.

\subsubsection{External Validity}
The evaluation case-study dataset used for evaluation was derived from RSE literature rather than from real-time practitioner queries or field deployments involving RSEs. This choice may bias the review toward models and metadata that are already well documented in research contexts, potentially inflating apparent system performance relative to scenarios involving sparsely documented or emerging models. Time constraints also prevented conducting longitudinal user studies or field evaluations, which could have provided more substantial evidence of the framework's practical utility in live workflows.

\smallskip

Additionally, the current implementation focuses exclusively on the Python/PyPI ecosystem. While \textsc{ModelSelect}'s architecture is language-agnostic and theoretically extendable to environments such as R, Julia, and Java, as well as to proprietary AI model registries, this has not yet been empirically validated. This could limit generalizability across languages, platforms, or industrial settings. Nonetheless, Python remains the most widely adopted ecosystem for AI development in RSE contexts, making it a pragmatic choice for initial instantiation.

\subsubsection{Reliability} 
Rationale annotation derived from publications inherently carries a degree of subjectivity. Even with clearly defined annotation guidelines, individual annotators may differ in their interpretation of what constitutes a rationale, particularly when rationales are implied rather than explicitly stated. While inter-annotator agreement was monitored, achieving consistency across all cases remains challenging. Future work could address this by incorporating larger annotator pools and adjudication procedures, as well as by developing automated heuristics for rationale extraction trained and validated against consensus-labeled data.

\smallskip

In addition, reliance on LLMs introduces temporal variability, as these models evolve. Changes in their architectures, training data, or default prompting strategies may lead to variations in outputs across runs, potentially affecting reproducibility. Although self-consistency checks and partial manual validation help mitigate this risk, full reproducibility cannot be guaranteed without access to the exact LLM model state at the time of execution, which requires version-controlled access. A practical mitigation strategy employed in this study involved using static random seeds and archiving intermediate outputs to establish reproducible inference environments.

\section{Related Work}
\label{related_work}

This section provides an overview of existing work in software engineering practices for AI development and AI model selection, and the broader problem of technology and commercial-off-the-shelf (COTS) selection, emphasizing the contributions of this study relative to prior approaches. Table \ref{table:literatureStudy} summarizes the methodologies used in the literature, categorized by decision-making domain, data collection method, weighting technique, and ranking method. Previous works are classified based on their similarity to the current study (methods used and sophistication) and are sorted by time in descending order. The bottom row indicates the number of prior works for each column, highlighting gaps in the literature.

\smallskip

The current study builds upon prior work by addressing limitations in scope, data collection, and ranking strategies, while introducing novel methods and a scalable framework for automated model selection.

\begin{table*}[t!]
    \centering
    \includegraphics[width=\textwidth,keepaspectratio]{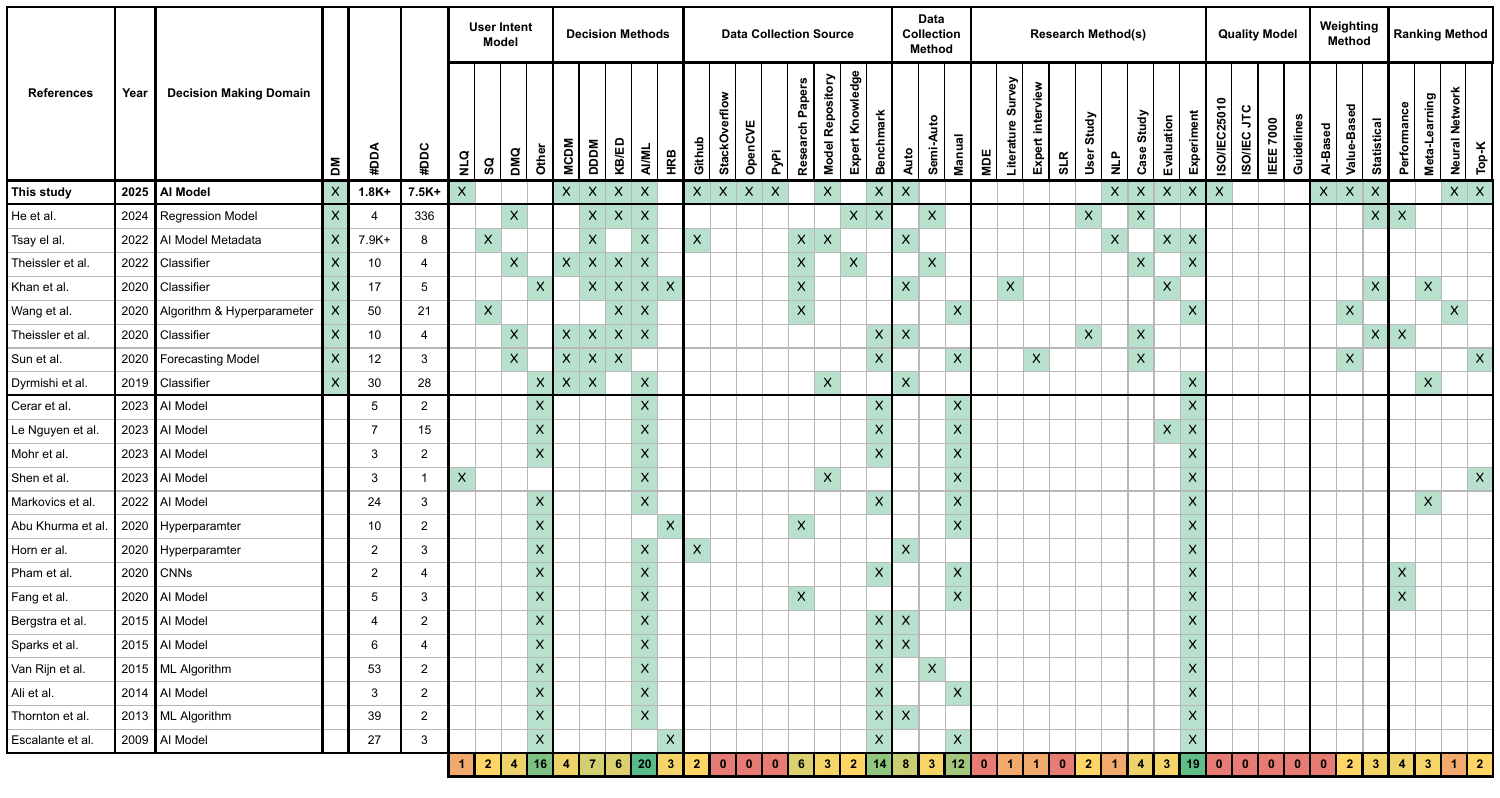}
    \caption{Summarization of methodologies used in AI model selection research, highlighting decision-making domains, data collection sources and methods, weighting techniques, and ranking strategies.}
    \label{table:literatureStudy}
\end{table*}

\subsection{Decision Making Domains}


Model selection strategies have been proposed for various applications, such as regression \cite{he2024vmsregression}, classification \cite{theissler2022, khan2020classifierselection, dyrmishi2019classifier}, forecasting \cite{sun2020forecasting}, and hyperparameter optimization \cite{wang2020automodel}. However, most earlier studies often focus on narrow domains with limited decision alternatives (e.g., a few models). 

\smallskip

Unlike previous approaches, this study expands the scope to general AI model selection, encompassing over 1,800 models and 7,500 features. These advancements address gaps in prior work, such as generalizability, providing a comprehensive tool for AI model selection that supports RSEs in making informed decisions.

\subsection{User Intent Modeling}

User intent modeling has gained attention in recent studies, particularly in the context of decision support systems. Prior work in this area often relies on decision-making queries (DMQ) \cite{tsay2022extracting, theissler2022, sun2020forecasting}, which include pairwise comparisons and prioritization schemes. These methods are typically used in systems that compare solutions based on multiple criteria, relying on explicit user preferences structured around predefined attributes. 

\smallskip

Another approach is structured queries (SQ) \cite{sun2020forecasting, wang2020automodel}, which use formal criteria and constraints to capture user intent, similar to SQL. Although these methods provide precision, they often suffer from limited flexibility and accessibility, as technical proficiency is required to use them.

\smallskip

In contrast, this study utilizes natural language querying (NLQ) similar to \cite{shen2023hugginggpt}, allowing users to express their intent in plain language. This approach significantly improves accessibility for non-experts and aligns with real-world use cases where structured queries may not be practical. By integrating NLQ with automated ranking and weighting techniques, the framework enhances user experience and decision accuracy, addressing a critical gap in the literature.

\subsection{Data Collection Sources and Methods}

Data collection is a cornerstone of AI model selection research, with prior studies mostly relying on diverse sources such as research papers \cite{tsay2022extracting},  \cite{theissler2022}, \cite{khan2020classifierselection}, \cite{wang2020automodel}, expert knowledge \cite{he2024vmsregression}, \cite{dyrmishi2019classifier}, model repositories \cite{tsay2022extracting}, and benchmarks \cite{sun2020forecasting}.

\smallskip

Most earlier approaches employ manual methods \cite{lenguyen2023comparative, khuma2020evolopyfs, cerar2023AIaaS, mohr2023fastandinformative, shen2023hugginggpt, markovics2022comparison}, utilizing the authors' expert knowledge for their model and dataset pools, and some employ semi-automatic methods \cite{he2024vmsregression, theissler2022}. These methods limit scalability and the ability to keep datasets up to date, and they rely on expert knowledge.

\smallskip

This study employs a fully automated data collection framework that aggregates information from GitHub repositories, Stack Overflow discussions, Open CVE, and library documentation. The inclusion of library documentation (AI library homepages and their complete website structure) makes a crucial difference by allowing \textsc{ModelSelect} to discover model repositories, benchmark datasets, and leaderboards indirectly, without needing to program specific crawlers for each source.



\subsection{Decision Alternatives and Criteria}

Prior studies often define decision alternatives based on the author's expertise or limited benchmarks, which can introduce bias and restrict the scope of model selection. For example, \cite{dyrmishi2019classifier} and \cite{theissler2022} evaluate classifiers using predefined criteria which does not account for the dynamic nature of AI development or the diverse needs of RSEs.

\smallskip

This study addresses these issues by incorporating a broader set of decision alternatives and criteria, and by leveraging automated data collection and GenAIs to process large volumes of textual data accurately. The integration of GenAIs in various pipelines replaces previously inaccurate NLP algorithms and reduces the manual effort required for these tasks. This enables the inclusion of a significant number of decision alternatives and criteria.

\subsection{Quality Models, Weighting, and Ranking Techniques}

Different studies in the literature employ various quality models. For instance, studies like \cite{serban2024swpractices} rely on custom evaluation guidelines. Similarly, \cite{deHond2022ai_prediction_guidelines} adopts \textit{ISO/IEC JTC 1/SC 42} \cite{iso2023sc42} and \textit{IEEE-7000-20} \cite{IEEE7000-2021}, which may not generalize to AI model selection tasks. None of these works, however, is related to the AI domain.

\smallskip

This work, as mentioned earlier, utilizes \textit{ISO/IEC 25010} \cite{ISO_IEC_25010_2023}, which can effectively evaluate the quality requirements of AI libraries compared to other standards.

\smallskip

Weighting and ranking are crucial components of model selection when presenting results to the user. Weighting involves evaluating the relevance of entries in the knowledge base based on the user's query and preferences. Prior studies, such as \cite{khan2020classifierselection} and \cite{theissler2022}, employ statistical metrics to assign weights, often using metrics such as accuracy or performance benchmarks. However, these approaches usually lack flexibility and fail to account for user-specific preferences or contextual factors.

\smallskip

This study introduces a hybrid approach that combines AI-based weighting, value-based scoring, and advanced ranking techniques. These methods enable dynamic adjustments based on user intent and model performance, ensuring that recommendations align with specific use cases. \cite{wang2020automodel} and \cite{sun2020forecasting} use value-based scoring that incorporates predefined priorities, such as favoring models with fewer vulnerabilities or greater popularity, based on collected metadata. Statistical metrics, such as those used by \cite{wang2020automodel}, are employed to refine weights based on measurable attributes, including hyperparameter optimization success rates.

\smallskip

Ranking methods further refine the selection by ordering models based on their weighted scores and limiting the number of options shown to the user. This study employs advanced ranking techniques, including Top-$k$ selection (similar to \cite{sun2020forecasting}) and neural network-based ranking. Top-$k$ selection is beneficial for scalability, as it filters the most relevant models from a large candidate pool. Performance methods \cite{pham2020skincancer} sort recommendations based on metrics such as accuracy in cases where benchmarks are available. Meta-learning (as in \cite{markovics2022comparison}) and neural network-based ranking methods \cite{wang2020automodel} enhance adaptability by learning patterns in user preferences or by adjusting the ordering of results using neural networks.

\section{Conclusion and Future Work}
\label{conclusion}

In this study, we introduced a data-driven decision framework for AI model selection based on MCDM principles. The approach integrates model features with quality measures using four automated data pipelines, a structured knowledge base, and an inference model to deliver context-aware, traceable recommendations. By combining model features and sentiment-based quality indicators within the MCDM paradigm, \textsc{ModelSelect} addresses key shortcomings of current AI model selection methods, including limited transparency, reproducibility, and quality-awareness.

\smallskip

Applied to a curated dataset of real-world case studies from diverse AI domains, \textsc{ModelSelect} achieved high coverage and strong rationale alignment in both model and library recommendation tasks. Pipeline-level experiments also demonstrated high precision in metadata extraction, model identification, and quality-assessment mapping, supporting the framework's scalability and reliability. These findings suggest that structured, empirically evaluated recommendation systems can help RSEs navigate the growing landscape of AI models.

\medskip

For future work, several directions can be explored to extend and strengthen the framework. First, while the current implementation focuses on the Python ecosystem, future work will explore other programming environments and model registries (e.g., R, Java, HuggingFace Hub) to test applicability across ecosystems. Second, building on the literature-based evaluation, we aim to conduct longitudinal user studies and controlled experiments with RSEs to assess the framework's impact on decision quality, efficiency, and trust in practice. Third, adding periodic re-execution of data pipelines and automated detection of new models, features, and rationales will help keep the knowledge base up to date. Finally, integrating a live user feedback loop could allow the ranking process to adapt to observed preferences, ensuring recommendations remain relevant and aligned with real-world needs.

\section*{Acknowledgments}

This research was conducted within the AI4RSE Laboratory of the Information Technology Group at Wageningen University \& Research, in close collaboration with the Department of Computer Science and Engineering at Shiraz University, as part of a collaborative research effort on evidence-driven approaches to AI for research software engineering.

The authors gratefully acknowledge the anonymous reviewers for their insightful comments and constructive suggestions, which significantly improved the quality of this manuscript. The authors also thank their colleagues for their valuable discussions and feedback throughout the course of this study.

\balance

\bibliographystyle{elsarticle-num-names}
\bibliography{references}












\end{document}